\documentclass[twocolumn,prb, amsmath,amssymb,letterpaper]{revtex4-1}

\usepackage{amsmath}
\usepackage{amssymb}
\usepackage{amsfonts}
\usepackage{graphicx}
\usepackage{dcolumn}
\usepackage{float}
\usepackage{bm}
\usepackage{mathrsfs}
\usepackage{txfonts}
\usepackage{tabularx}
\usepackage{bigstrut}
\usepackage[usenames,dvipsnames]{color}
\usepackage{verbatim}

\begin{document}

\title{Strong and Anisotropic Third Harmonic Generation in Monolayer and Multilayer ReS$_2$ }

\author{Qiannan Cui$^{1}$}
\author{Rodrigo A. Muniz$^{2}$}
\author{J. E. Sipe$^{2}$}
\author{Hui Zhao$^{1}$}
\email{huizhao@ku.edu}

\affiliation{$^{1}$Department of Physics and Astronomy, The University of Kansas, Lawrence, Kansas 66045, United States}

\affiliation{$^{2}$Department of Physics and Institute for Optical Sciences, University of Toronto, Toronto, Ontario MS5 1A7, Canada}

\date{\today}

\begin{abstract}
We report observation of strong and anisotropic third harmonic generation (THG) in monolayer and multilayer ReS$_2$. 
The third-order nonlinear optical susceptibility of monolayer ReS$_2$,  $\left| \chi^{(3)} \right|$ is on the order of $10^{-18} $ m$^2$/V$^2$, which is about one order of magnitude higher than reported results for hexagonal-lattice  transition metal dichalcogenides such as MoS$_2$. 
A similar magnitude for the third-order nonlinear optical susceptibility was also obtained for a multilayer sample. 
The intensity of the THG field was found to be dependent on the direction of the incident light polarization for both monolayer and multilayer samples. 
A point group symmetry analysis shows that such  anisotropy is not expected from a perfect $1T$ lattice, and must arise from the distortions in the ReS$_2$ lattice.  
Our results show that THG measurements can be used to characterize lattice distortions of two-dimensional materials, and that lattice distortions are important for the nonlinear optical properties of such materials. 
\end{abstract}

\maketitle

\section{Introduction}

Transition metal dichalcogenides (TMDs) constitute a new generation of semiconducting materials. 
Their electronic and optical properties have been extensively studied since the discovery of the direct band gap in monolayer MoS$_2$ in 2010 \cite{mos2di,feng}, as understanding light-matter interactions in TMD monolayers is essential for developing applications in electronics, photonics, and optoelectronics. 
So far, ultrafast excitonic dynamics \cite{ultrafast1}, coupled spin-valley physics \cite{spinvalley,muniz}, large exciton binding energies \cite{binding1, binding2}, and nonlinear optical responses \cite{shgm, thgm, thgw,sat} in TMD monolayers have been well studied. 
It has been also shown that van der Waals heterostructures with different TMD monolayers can be constructed \cite{van} to realize devices with multiple functionalities.

After much effort, ReS$_2$ has been identified as an anomalous member of the TMD family.
In 2014, few-layer and bulk samples of ReS$_2$ were reported to display monolayer behavior \cite{mono1}. 
Since then the optical and electrical properties of ReS$_2$ have been intensively investigated.
Coherent control of ballistic transport has been realized in bulk ReS$_2$ \cite{ballistic}. 
The in-plane anisotropic optical and electrical properties of monolayer ReS$_2$ have also been experimentally studied \cite{ani2,ani1}. 
A recent work has shown that  stacking orders can be resolved by Raman spectroscopy, due to the unique crystal structure of ReS$_2$ \cite{stacking}. 
Based on these novel properties, a wide spectrum of applications using ReS$_2$ has been proposed. 
Among them are energy storage devices \cite{ap2, ap3}, integrated digital converters \cite{ap7}, and sensitive photodetectors \cite{ap1,ap4,ap5,ap6}. 
Thus ReS$_2$ has not only become a unique platform for novel 2D physics, but also an unusual member of TMDs to construct van der Waals heterostructures.

Unlike the TMD monolayers based on Mo and W, which form in a hexagonal lattice, monolayer ReS$_2$ has a stable distorted $1T$ crystal lattice \cite{ReS2s}. 
Re atomic chains formed by Re-Re bonding run along the direction of the $b$-axis, enabling the in-plane anisotropy of electrical and optical response. 
However, to date there have been no reports of the nonlinear optical properties of ReS$_2$. 
Since nonlinear optical properties are known to be extremely sensitive to the lattice symmetries  \cite{sipe1,sipe2}, it is important to determine to what extent the lattice distortions of ReS$_2$ affect its nonlinear optical properties.

In this paper, we report experimental results of third harmonic generation (THG) in monolayer and multilayer ReS$_2$. 
We determine the third-order susceptibility of monolayer ReS$_2$ and find that its magnitude is about one order of magnitude larger than those of hexagonal TMDs such as MoS$_2$ \cite{thgm}.
We also find that the third-order nonlinear response has strong in-plane anisotropies incompatible with an undistorted $1T$ lattice.

The outline of the paper is the following: In Sec. \ref{sec:setup} we describe the experimental setup and samples, and in Sec. \ref{sec:theory} we discuss  the expected features in the results for materials with a perfect $1T$ lattice, and in Sec. \ref{sec:results} we show the experimental results obtained for ReS$_2$. 
We present the conclusions of this study in Sec. \ref{sec:conclusion}. 

\section{Experimental Setup and Samples}
\label{sec:setup}

\begin{figure}[tb!]
  \centering
  \includegraphics[width=0.85\columnwidth]{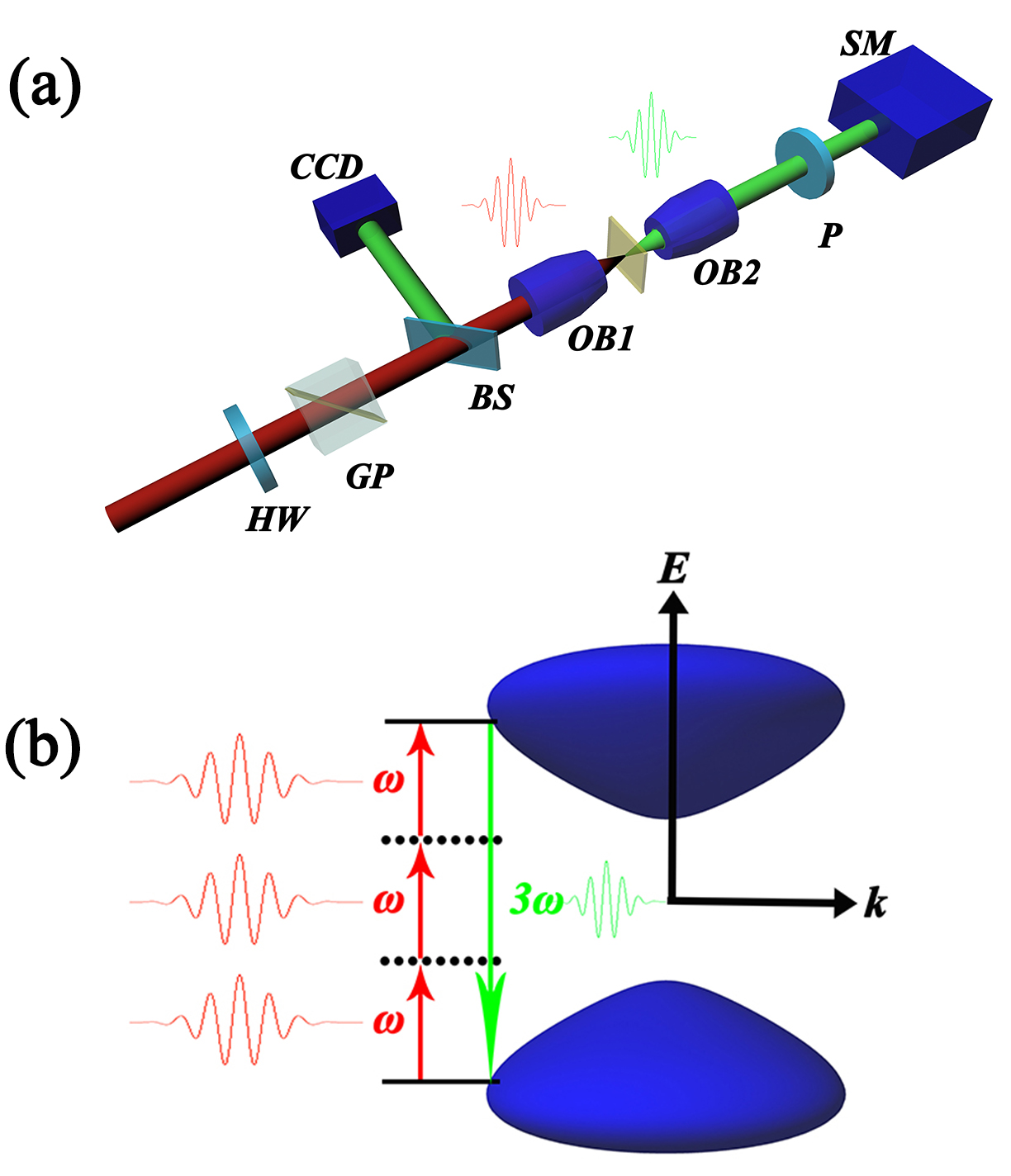}
  \caption{(a) Experimental setup: the infrared femtosecond laser (red pulse) is tightly focused on the sample by an objective lens (OB1). The induced THG signal (green pulse) is directly measured by a spectrometer(SM). HW is a half wave plate, GP represents a Glan prism, and P is a polarizer. (b) The diagram illustrates the THG due to the fundamental pulse $\omega$ and the monolayer ReS$_2$ bandstructure.}
      \label{F1}
\end{figure}

We use $250$ fs pulses generated from an optical parametric oscillator pumped by a Ti:sapphire near-infrared laser.
As shown in Figure \ref{F1}(a), a half wave plate (HW) and a Glan prism (GP) are used to adjust the power and the polarization direction of the fundamental pulses. The fundamental pulses at $\omega$ are tightly focused on the sample by an objective lens (OB1) with a numerical aperture (NA) of 0.42. The generated third harmonic pulses are collected by another objective lens (OB2) with the same NA. 
A spectrometer (SM) is employed to directly measure the power of the third harmonic pulses at $3\omega$, and a polarizer (P) before the spectrometer is used to resolve the horizontal and vertical components of the third harmonic pulses. 
A charge-coupled device (CCD), a beamsplitter (BS), and the OB1 serve as a microscope to monitor the focusing position on the sample by collecting the reflected light at $3\omega$. 

Figure \ref{F1}(b) shows a diagram of THG in monolayer ReS$_2$ with a simplified band structure, where three photons at $\omega$ generate one photon at $3\omega$. 
The fundamental photon energy is about $\hbar\omega =0.82$ eV (1515 nm), while the energy of the generated photons is about $3\hbar\omega =2.46$ eV (505 nm). Since the band gap of monolayer ReS$_2$ is about 1.53 eV (810 nm), the fundamental pulse can not be absorbed by monolayer ReS$_2$ via one-photon absorption. 
The ReS$_2$ samples are fabricated on the surface of PDMS by mechanical exfoliation and transferred to a BK7 glass substrate about 0.48 mm thick \cite{dry}. 
All measurements are carried out in ambient conditions.

\begin{figure}[h]
  \centering
  \includegraphics[width=1\columnwidth]{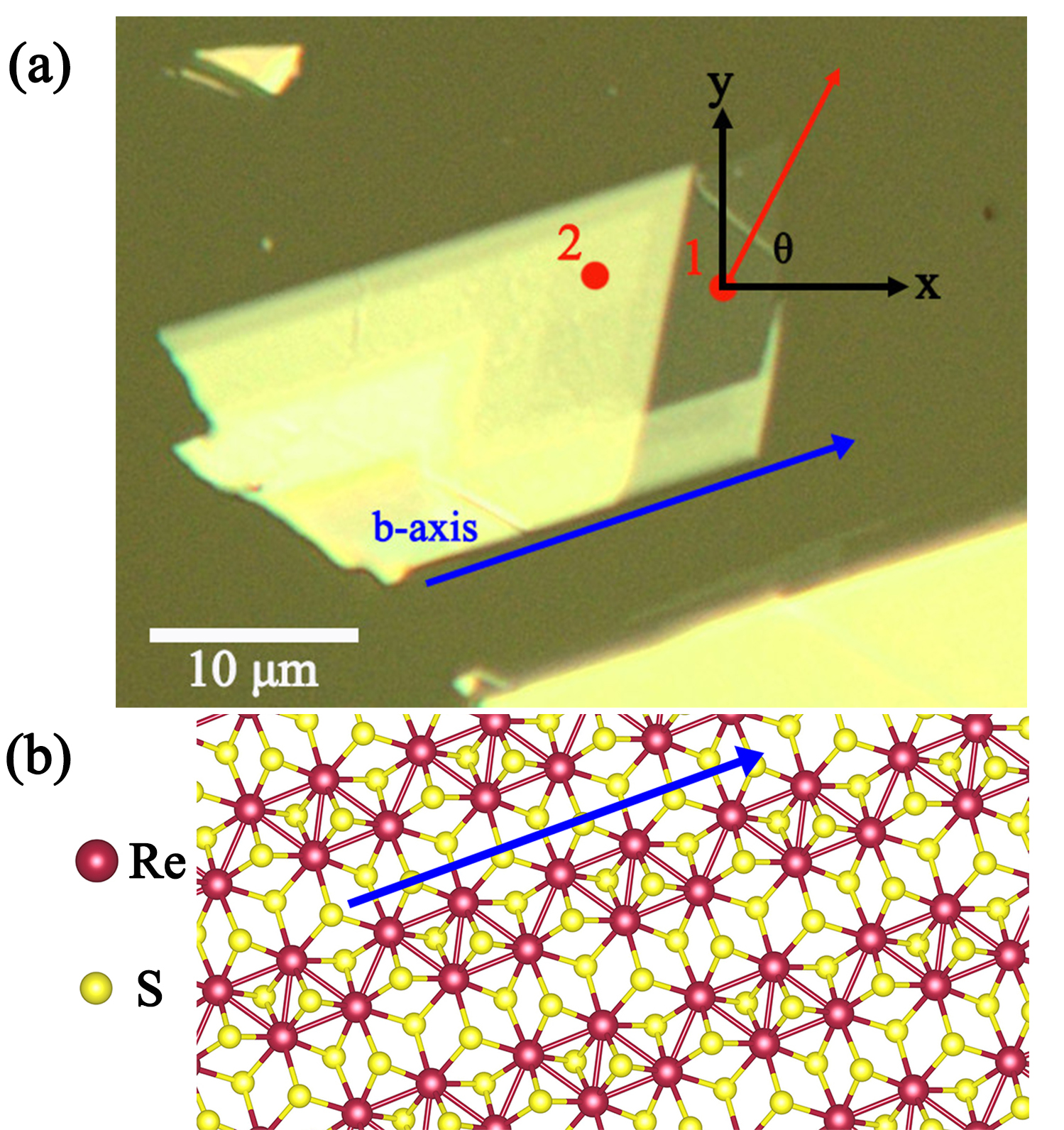}
\caption{(a) Microscope image of monolayer ReS$_2$ and attached multilayer ReS$_2$. The red dots 1 and 2 are the laser focusing positions for monolayer and multilayer measurements respectively. The blue arrow indicates the $b$-axis direction, which is the direction of the Re atom chains in the lattice. The $x$ and $y$ axes define the lab coordinates, and $\theta$ denotes the angle between the polarization direction of the fundamental pulse (red arrow) and the $x$-axis. (b) The distorted $1T$ lattice structure of ReS$_2$, showing the Re atom chain along the $b$-axis (blue arrow).}
      \label{F2}
\end{figure}

An optical microscope image of the monolayer ReS$_2$ sample is shown in Figure \ref{F2}(a), where the monolayer flake is attached to a multilayer. 
The red dots 1 and 2 in the same figure indicate the focused fundamental laser spots on the monolayer and multilayer, respectively. 
The figure also shows laboratory axes $x$ and $y$, as well as the angle $\theta$ between the fundamental pulse polarization and the $x$-axis. 
The lattice structure of ReS$_2$ monolayer is depicted in Figure \ref{F2}(b);  
Re atom chains are formed along the $b$-axis direction, which is depicted by the blue arrow and experimentally determined by transient absorption measurement (see Supplemental Material). As shown in Figure \ref{F2}(a), the $b$-axis direction is along one of the two cracked edges of the flake.

\begin{figure}[h]
  \centering
  \includegraphics[width=1\columnwidth]{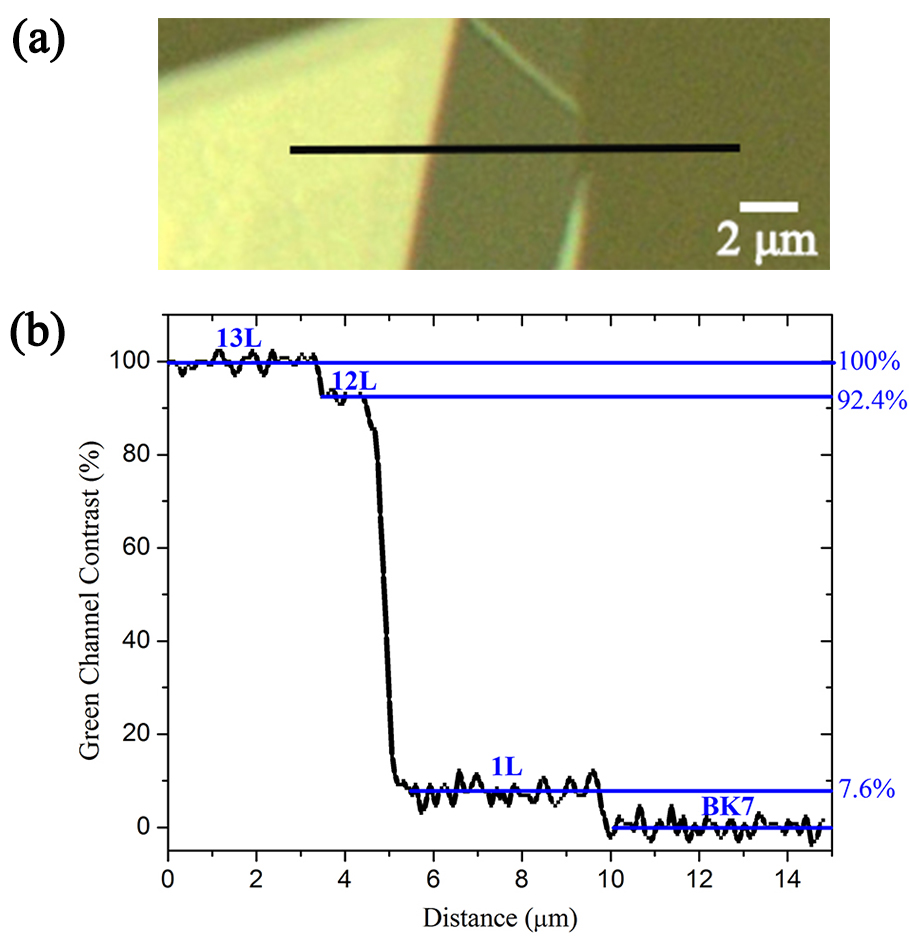}
  \caption{(a) Zoom in of the sample region used for optical contrast calculations. (b) Normalized optical contrast of green channel along the black line indicated in (a). }
      \label{F3}
\end{figure}

An optical contrast technique \cite{tony} is employed to determine the thickness of the samples. 
In Figure \ref{F3}(a), the black line indicates a cross section of our monolayer and multilayer samples. 
As shown in Figure \ref{F3}(b), the normalized optical contrast of the monolayer (1 L, 0.73 nm thick) and multilayer samples in the green channel are about 7.6\ \% and 100\ \%, respectively. 
Since each layer increases the optical contrast by roughly 7.6\ \%, the multilayer sample is identified as 13 L, which is about 9.5 nm thick. 
Similarly, the transition region with an optical contrast of $92.4\ \%$ is about 1 $\mu$m wide and identified as 12 L.

\section{Point group symmetry analysis of THG in $1T$ layered materials }
\label{sec:theory}

In this section we carry out a point group symmetry analysis for the THG of undistorted $1T$ layered materials; we use the lab coordinates shown in Figure \ref{F2}(a). 

The polarization associated with THG is determined by the third order susceptibility $\chi^{(3)}_{abcd}\left(\omega,\omega,\omega\right)$, which is a rank-4 tensor that respects the symmetries of the point group of the lattice,  
and the applied electric field ${\bf E}\left(t\right)={\bf E}\left(\omega\right)e^{-i\omega t} +c.c.$,
through 
\begin{equation}
P_{a}\left(3\omega\right) = \epsilon_0 \chi^{(3)}_{abcd}\left(\omega,\omega,\omega\right)E_{b}\left(\omega\right)E_{c}\left(\omega\right)E_{d}\left(\omega\right).
\end{equation}
The electric field associated with the generated radiation is proportional to the polarization 
${\bf P}\left(3\omega\right)$, so the intensity of
the generated radiation is proportional to 
$\left|P\left(3\omega\right)\right|^{2}$. 

\begin{figure}[h]
  \centering
  \includegraphics[width=0.85\columnwidth]{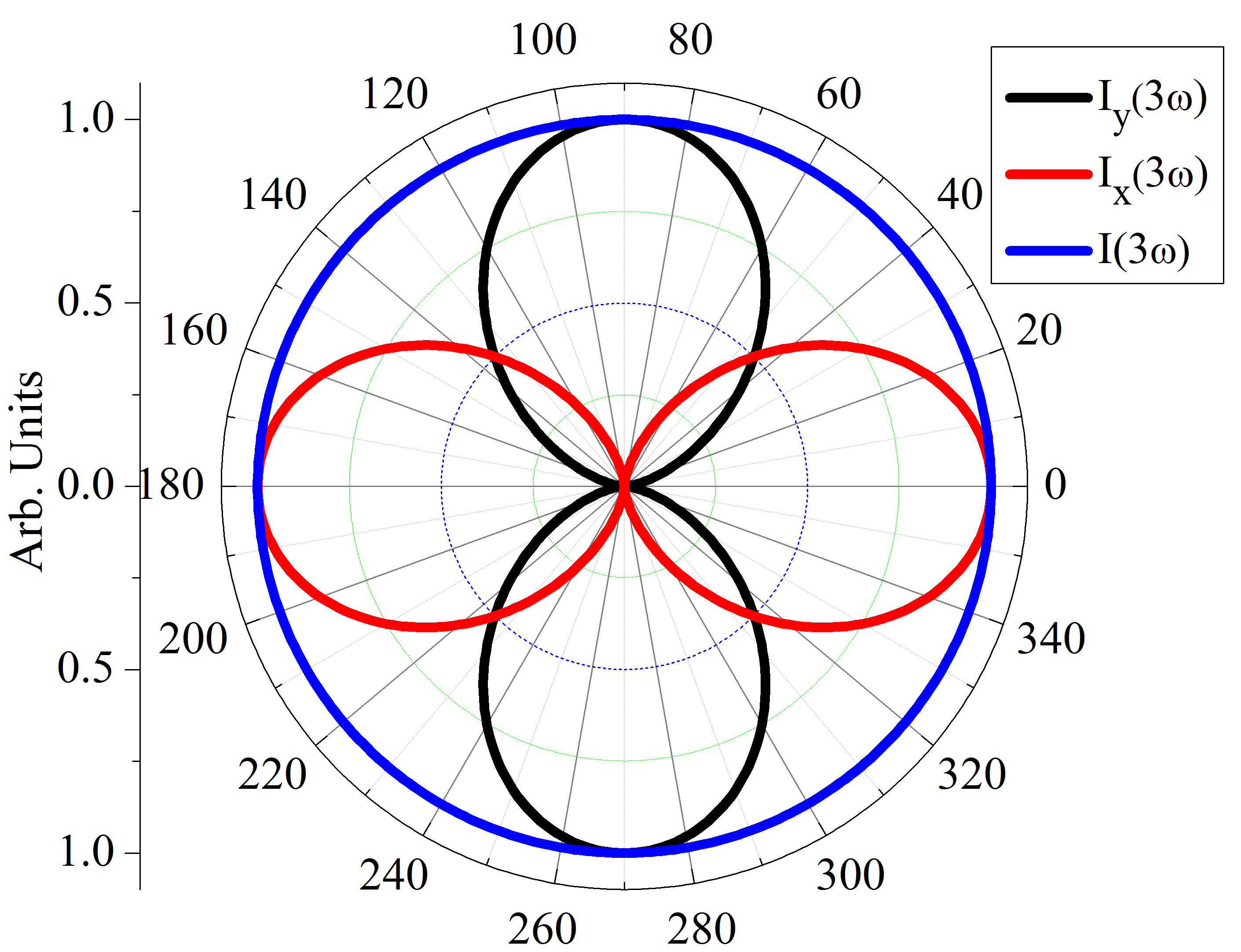}
  \caption{The intensities of the different polarizations of the THG field are shown as a function of the angle of polarization of the incident field.
The red line corresponds to the horizontal ($x$-axis) polarization, the black line corresponds to vertical ($y$-axis) polarization, and the blue line corresponds to the total intensity of the THG field.}
      \label{F4}
\end{figure}

The point group of an undistorted $1T$ lattice  is $D_{3d}$ ($\bar{3}\frac{2}{m}$), but in the presence of a substrate the symmetry operations that act on the perpendicular axis are broken, and the point group is $C_{3v}$ ($3m$). 
Only the planar components of $\chi^{(3)}_{abcd}$ are relevant for normal incidence, and there are only 3 such independent components for both point groups $D_{3d}$ and $C_{3v}$, namely $\chi_{xxxx}^{(3)}$, $\chi_{xxyy}^{(3)}$ and $\chi_{xyyx}^{(3)}$. 
Consequently, the THG has the same angular dependence regardless of the presence of the substrate. 
Describing the incident field as
${\bf E}\left(\omega\right)=E\left(\omega\right) \left( \hat{\bf x}\cos\theta+\hat{\bf y}\sin\theta \right)$, the polarization associated with the THG field is given by 
\begin{equation}
\begin{array}{rl}
P_{x}\left(3\omega\right)= & \epsilon_0 \chi_{xxxx}^{(3)} \left[ E\left(\omega\right) \right]^{3} \cos\left(\theta\right), \\
P_{y}\left(3\omega\right)= & \epsilon_0 \chi_{xxxx}^{(3)} \left[ E\left(\omega\right) \right]^{3} \sin\left(\theta\right),
\end{array}
\label{polariz}
\end{equation}
and the intensities of the THG fields horizontally and vertically polarized are 
\begin{equation}
\begin{array}{rl}
I_{x}\left(3\omega\right)= &  \frac{1}{2}I_{max}\left[1+\cos\left(2\theta\right)\right],\\
I_{y}\left(3\omega\right)= &  \frac{1}{2}I_{max}\left[1-\cos\left(2\theta\right)\right].
\end{array}
\end{equation}
Thus the total intensity $I_{x} +I_{y}$ is independent of the polarization direction of the incident fields.
The above expressions are plotted in Figure \ref{F4}, which shows the isotropy of the total intensity of the THG from an undistorted $1T$ lattice. 
Thus any anisotropy in the experimental results of the total THG for ReS$_2$ is due to its lattice distortions. 

\section{Experimental Results and Discussion}
\label{sec:results}

\begin{figure}[ht!]
  \centering
  \includegraphics[width=0.9 \columnwidth]{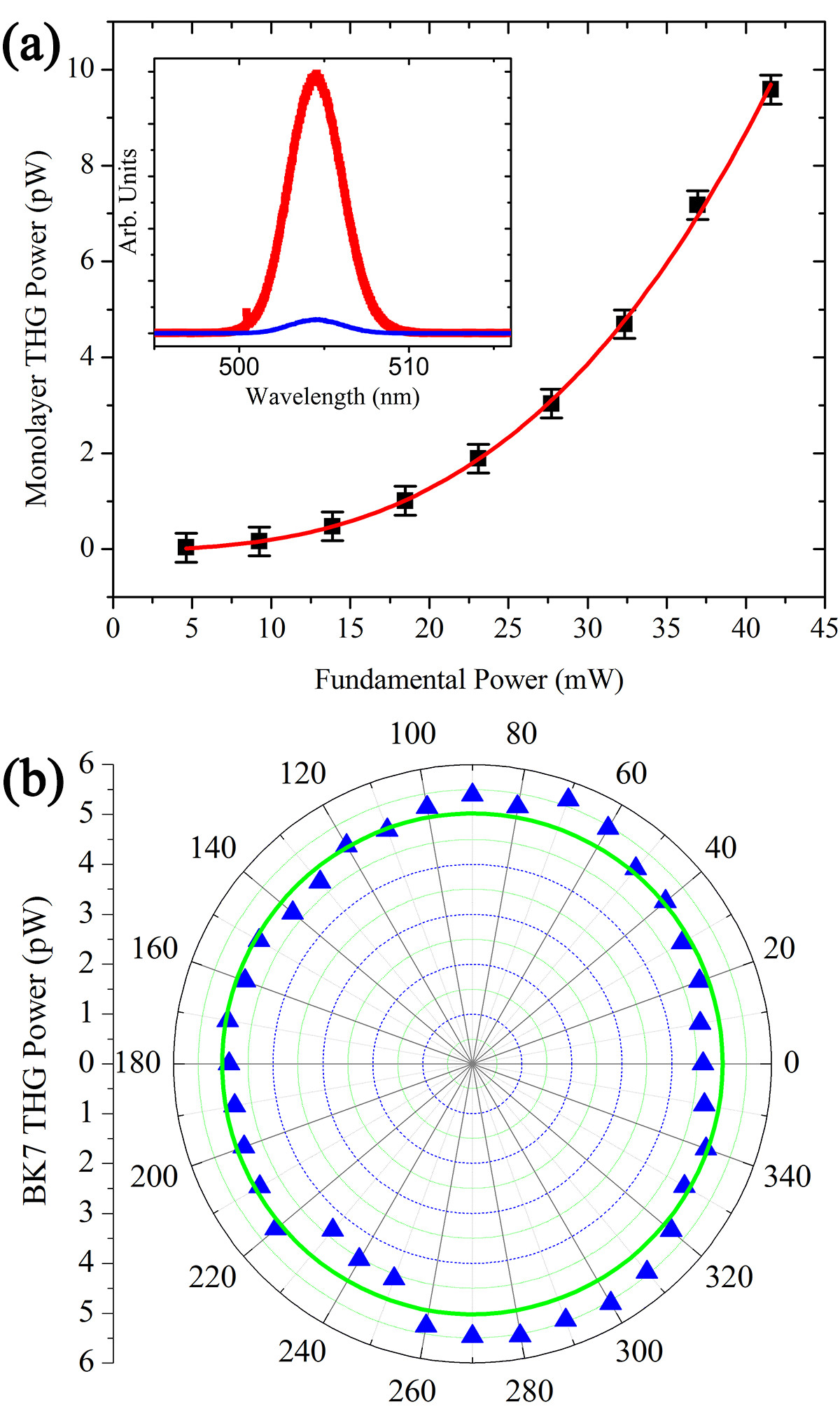}
  \caption{(a) THG power (black rectangles) of monolayer ReS$_2$ as a function of fundamental power. The red line is a cubic fitting. The inset illustrates the third harmonic spectrum from the monolayer ReS$_2$ sample (red) and the BK7 glass substrate (blue) under the same experimental conditions. (b) Angular dependence of THG (blue triangles) in BK7 glass substrate when the fundamental pulse power is fixed at 23 mW. The green line is an isotropic fitting. }
      \label{F5}
\end{figure}

We study THG of monolayer ReS$_2$ by focusing the fundamental field on the red dot 1 as shown in Figure \ref{F2}(a). 
The THG spectrum is shown as the red curve in the inset of Figure \ref{F5}(a), which corresponds to $\theta =310^\circ$ and a fundamental fluence of 12.6 mJ/cm$^2$. 
In our experiments, third harmonic frequencies can also be generated from the BK7 glass substrate. 
The blue curve in the inset of Figure \ref{F5}(a), which is about $5\ \%$ of the red curve, shows the third harmonic spectrum of the BK7 glass substrate under the same experimental conditions as the monolayer ReS$_2$ sample. 
In addition, we notice that THG from BK7 glass substrate becomes maximal when moving the $\omega$ beam focus away from the surface and into the BK7 glass substrate. 
Thus we study THG from the BK7 glass substrate by moving the focus of the fundamental beam into the substrate by a few $\mu$m. 
There is no angular dependence of THG power from the  BK7 glass substrate when the power of the fundamental beam is fixed at 23 mW, as shown in  Figure \ref{F5}(b), and the contribution from the BK7 glass substrate is much smaller than the THG from ReS$_2$ samples; it has no effect in our analysis. 
The fundamental power dependence of THG in monolayer ReS$_2$ is shown in Figure \ref{F5}(a) for $\theta = 310^\circ$. The cubic power fit (red line) matches well with the data (black rectangles), as expected for a THG process. 

\begin{figure}[h]
  \centering
  \includegraphics[width=1\columnwidth]{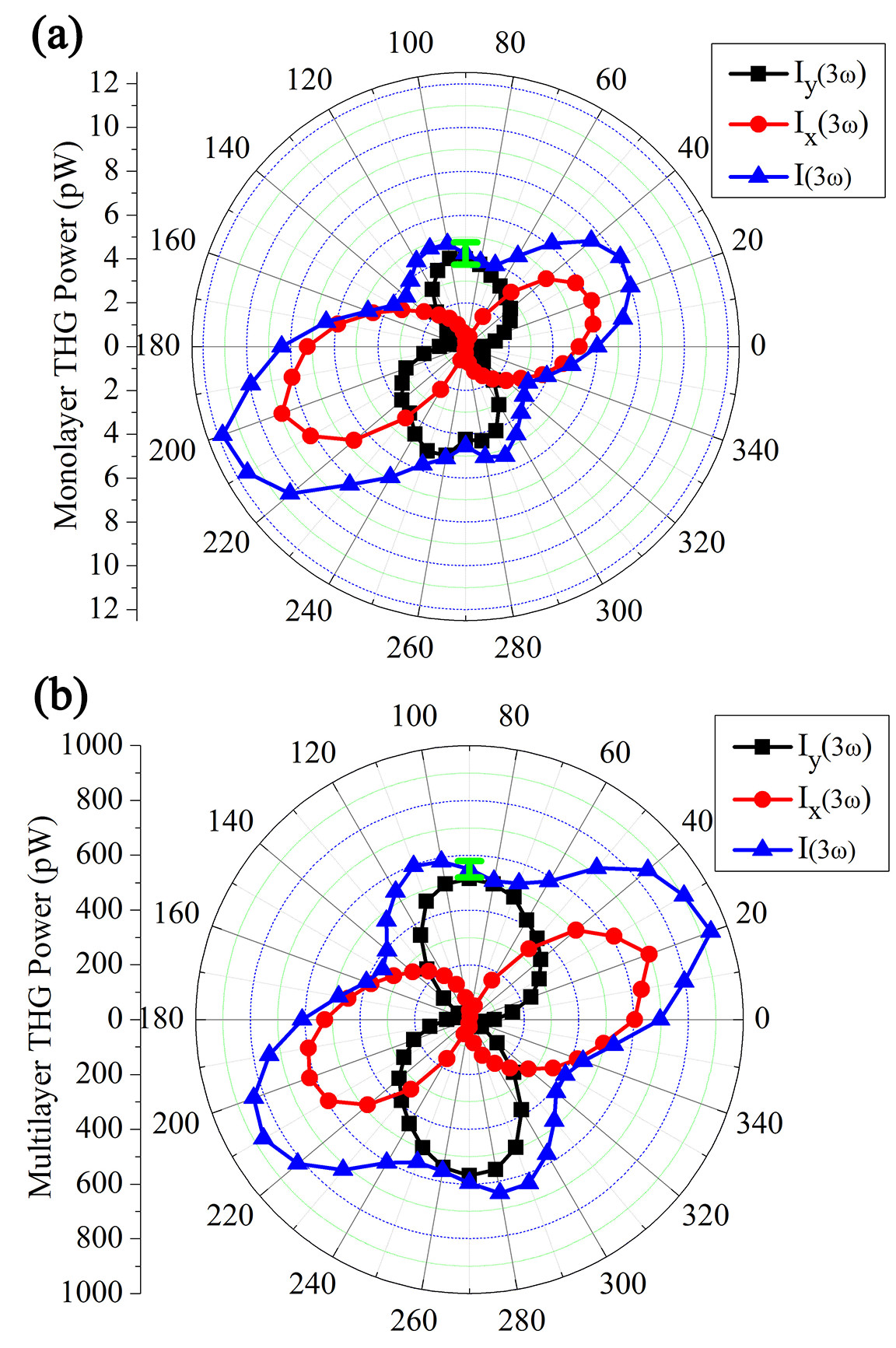}
  \caption{Angular dependence of THG power in monolayer ReS$_2$ (a) and attached multilayer ReS$_2$ (b). Horizontal, vertical components of total THG power (blue) are in red and black, respectively. The fundamental pulse power is fixed at 30.2 mW. }
      \label{F6}
\end{figure}

By tuning the HW and the GP we measure THG in monolayer ReS$_2$ as a function of $\theta$, under the same fundamental fluence of 16.5 mJ/cm$^2$. Horizontal and vertical polarizations of the THG field are resolved by the polarizer before the spectrometer. 
The results are plotted in Figure \ref{F6}(a), where the lines corresponding to the horizontal and vertical polarizations have the shape of a twisted dumbbell, and the total power (i.e. the sum of the horizontal and vertical components) of the THG field from monolayer ReS$_2$ is anisotropic as a function of $\theta$.

In order to quantify the THG in monolayer ReS$_2$, we deduce the magnitude of the third-order susceptibility $\left| \chi^{(3)} \right|$ from the power of the THG field that reaches the detector $\bar{P}_{a}\left(3\omega \right)$ and  that of the fundamental incident field $\bar{P}_{in}\left(\omega\right)$.
Here we extract an estimate for the magnitude of the largest component of $\chi^{(3)}$, which we denote by $\left| \chi^{(3)} \right|$.
The fundamental field at the sample is related to the incident field by Fresnel equations. 
The induced polarization is related to the fundamental field at the sample by Eq. (\ref{polariz}).
The THG field at the sample can then be determined from the polarization \cite{sipe87}, which in turn determines the field that is transmitted through the glass substrate and collected by the detector. 
Considering pulses that are Gaussian in both space and time, the final expression is 
\begin{equation}
\left|\chi^{\left(3\right)}\right|=\left(\dfrac{\pi}{\ln2}\right)^{\frac{3}{2}} \dfrac{\left(n_{a}+n_{g}\right)^{5}}{8n_{a}^{2}n_{g}}\dfrac{c^2 \epsilon_{0}}{3\omega d}f_{rep}\dfrac{\tau}{2}\left(\dfrac{W}{2}\right)^{2}\sqrt{\dfrac{3\sqrt{3}\bar{P}_{a}\left(3\omega\right)}{\bar{P}_{in}\left(\omega\right)^{3}}},
\label{eq:chi3power}
\end{equation}
where $d$ is the thickness of the sample, $n_{a}=1$ and $n_{g}=1.521$ are respectively the indices of refraction of air and the BK7 glass substrate (which are both independent of the frequencies involved, $\omega$ and $3\omega$); $W$ and $\tau$ are respectively the spot size width and the
pulse duration measured at full width at half maximum, and $f_{rep}=80.54$ MHz is the laser  repetition rate.  
See the Supplemental Material \cite{supplement} for details on the derivation of this equation.  
 By using the parameters measured in our experiment ($d = 0.73$ nm, $\tau = 250$ fs, $W = 3.5$ $\mu$m,  and $\lambda = 1515$ nm), we extract the maximal value of $\left| \chi^{(3)}_{20^\circ} \right| \sim 5.3 \times 10^{-18}$ m$^2$/V$^2$ for $\theta = 20 ^\circ$.  
The maximal THG is confirmed to occur when the polarization of the fundamental field is along the $b$-axis, as determined by angular dependent transient absorption. 
Thus, THG measurements can be also used to determine the lattice orientation of ReS$_2$. 
 When $\theta = 70^\circ$, the THG field power reduces to about one third of the maximal value, so  $\left| \chi^{(3)}_{70^\circ} \right| \sim 3.5\times10^{-18}$ m$^2$/V$^2$ is the minimal value of the susceptibility.

To confirm the procedure used to extract the $\left| \chi^{(3)} \right|$ of the monolayer ReS$_2$ sample, we move the fundamental pulse focus spot to the BK7 glass substrate, which allows us to measure the THG from the BK7 glass substrate under the same conditions. 
The angular dependence of THG in the BK7 glass substrate is isotropic, as shown in Figure \ref{F5}(b). 
We extract a $\left| \chi^{(3)}_{BK7} \right| \sim2.27 \times10^{-22}$ m$^2$/V$^2$ for the BK7 substrate, which is in good agreement with the reported values of $\left| \chi^{(3)}_{BK7} \right| = 2.98 \times10^{-22}$ m$^2$/V$^2$ for a fundamental wavelength of 1064 nm and $\left| \chi^{(3)}_{BK7} \right| = 2.38 \times10^{-22}$ m$^2$/V$^2$ for 1907 nm \cite{BK7}.
The power of the THG due to the substrate is more than $20$ times smaller than that of the ReS$_2$, as shown in the inset of Fig. \ref{F5} (a). Thus, even the heterodyne contribution of the substrate to the total power of the THG is at least $10$ times smaller than that due exclusively to the ReS$_2$ sample, which lies within the experimental precision. So the contribution of THG from the BK7 glass substrate is ignored in our analysis. 

We also use the method described above to study the THG from multilayer ReS$_2$ by focusing the fundamental laser on the attached multilayer part of the sample, as indicated by the red dot 2 in Fig. \ref{F2}(a). 
The observed THG from the attached multilayer (13 L) shows an angular dependence very similar to that of monolayer ReS$_2$. 
This is a reasonable result, since the attached multilayer should have the same crystal orientation of the monolayer, resulting in the same in-plane symmetry of THG. 
In addition, the THG signal in the multilayer sample is significantly larger than that from the monolayer. 
Due to the weak interlayer coupling in ReS$_2$, the THG field from a multilayer sample can be considered as the addition of THG fields generated in each individual layer, and the THG induced power would be simply expected to be proportional to the square of the sample thickness $d$ -- see Eq. \ref{eq:chi3power}. 
Thus, the THG power from the 13 L sample would be expected to increase by a factor of 169, relative to the monolayer, for each $\theta$. 
However, since the photon energy of THG is higher than the band gap of ReS$_2$, the induced THG will be absorbed via one-photon absorption when propagating in the multilayer sample. 
It has been well-known that linear absorption in ReS$_2$ is in-plane anisotropic. A maximum (minimum) absorption coefficient is expected when light polarization is along (perpendicular to) the $b$-axis, which is clearly revealed by our experimental results: Fig. \ref{F6} shows that when compared to the monolayer THG, the multilayer THG is mostly suppressed by absorption when the polarization of the induced THG is along the $b$-axis ($\theta = 20^\circ$).

Finally, the distorted dumbbell shapes for the horizontal and vertical polarizations of the THG field are consistent with the predictions from the point group symmetry analysis. Since we have used BK7 glass substrate as a reference material, we rule out any anisotropic artifact from the measurements, and we can safely attribute the anisotropic THG to the lattice distortions of ReS$_2$.

\section{Conclusion}
\label{sec:conclusion}

In summary, we observed strong and anisotropic third harmonic generation (THG) in monolayer and multilayer ReS$_2$. The third-order susceptibility of monolayer ReS$_2$, $\left| \chi^{(3)} \right|$ was found to be on the order of 10$^{-18}$ m$^2$/V$^2$, which is about one order of magnitude higher than reported values in hexagonal TMDs, such as MoS$_2$. This large nonlinear optical response suggests potential applications of this material in optoelectronic devices involving third-order nonlinear processes. 
A point group symmetry analysis indicates that the THG in crystals with a perfect $1T$ lattice would be  isotropic, i.e. the total THG intensity would be  independent of the polarization of the incident field. 
However, we observed significant deviation from this prediction, which can be attributed to the lattice distortions. This illustrates the importance of lattice distortions for the nonlinear optical response of this material, and indicate that they should not be neglected in theoretical models. Our results also show the THG can be used as a diagnostic tool to probe such lattice distortion.

\section{Acknowledgement}

This material is based upon work supported by the National Science Foundation of USA under Award No. DMR-1505852 and IIA-1430493.

%\bibliographystyle{unsrt}

%\bibliography{ref}

\begin{thebibliography}{0}%
\makeatletter
\providecommand \@ifxundefined [1]{%
 \@ifx{#1\undefined}
}%
\providecommand \@ifnum [1]{%
 \ifnum #1\expandafter \@firstoftwo
 \else \expandafter \@secondoftwo
 \fi
}%
\providecommand \@ifx [1]{%
 \ifx #1\expandafter \@firstoftwo
 \else \expandafter \@secondoftwo
 \fi
}%
\providecommand \natexlab [1]{#1}%
\providecommand \enquote  [1]{``#1''}%
\providecommand \bibnamefont  [1]{#1}%
\providecommand \bibfnamefont [1]{#1}%
\providecommand \citenamefont [1]{#1}%
\providecommand \href@noop [0]{\@secondoftwo}%
\providecommand \href [0]{\begingroup \@sanitize@url \@href}%
\providecommand \@href[1]{\@@startlink{#1}\@@href}%
\providecommand \@@href[1]{\endgroup#1\@@endlink}%
\providecommand \@sanitize@url [0]{\catcode `\\12\catcode `\$12\catcode
  `\&12\catcode `\#12\catcode `\^12\catcode `\_12\catcode `\%12\relax}%
\providecommand \@@startlink[1]{}%
\providecommand \@@endlink[0]{}%
\providecommand \url  [0]{\begingroup\@sanitize@url \@url }%
\providecommand \@url [1]{\endgroup\@href {#1}{\urlprefix }}%
\providecommand \urlprefix  [0]{URL }%
\providecommand \Eprint [0]{\href }%
\providecommand \doibase [0]{http://dx.doi.org/}%
\providecommand \selectlanguage [0]{\@gobble}%
\providecommand \bibinfo  [0]{\@secondoftwo}%
\providecommand \bibfield  [0]{\@secondoftwo}%
\providecommand \translation [1]{[#1]}%
\providecommand \BibitemOpen [0]{}%
\providecommand \bibitemStop [0]{}%
\providecommand \bibitemNoStop [0]{.\EOS\space}%
\providecommand \EOS [0]{\spacefactor3000\relax}%
\providecommand \BibitemShut  [1]{\csname bibitem#1\endcsname}%
\let\auto@bib@innerbib\@empty
%</preamble>
\end{thebibliography}%


\begin{thebibliography}{99}
\expandafter\ifx\csname natexlab\endcsname\relax\def\natexlab#1{#1}\fi
\expandafter\ifx\csname bibnamefont\endcsname\relax
  \def\bibnamefont#1{#1}\fi
\expandafter\ifx\csname bibfnamefont\endcsname\relax
  \def\bibfnamefont#1{#1}\fi
\expandafter\ifx\csname citenamefont\endcsname\relax
  \def\citenamefont#1{#1}\fi
\expandafter\ifx\csname url\endcsname\relax
  \def\url#1{\texttt{#1}}\fi
\expandafter\ifx\csname urlprefix\endcsname\relax\def\urlprefix{URL }\fi
\providecommand{\bibinfo}[2]{#2}
\providecommand{\eprint}[2][]{\url{#2}}

\bibitem[{\citenamefont{Mak et~al.}(2010)\citenamefont{Mak, Lee, Hone, Shan,
  and Heinz}}]{mos2di}
\bibinfo{author}{\bibfnamefont{K.~F.} \bibnamefont{Mak}},
  \bibinfo{author}{\bibfnamefont{C.}~\bibnamefont{Lee}},
  \bibinfo{author}{\bibfnamefont{J.}~\bibnamefont{Hone}},
  \bibinfo{author}{\bibfnamefont{J.}~\bibnamefont{Shan}}, \bibnamefont{and}
  \bibinfo{author}{\bibfnamefont{T.~F.} \bibnamefont{Heinz}},
  \bibinfo{journal}{Phys. Rev. Lett.} \textbf{\bibinfo{volume}{105}},
  \bibinfo{pages}{136805} (\bibinfo{year}{2010}).

\bibitem[{\citenamefont{Splendiani et~al.}(2010)\citenamefont{Splendiani, Sun,
  Zhang, Li, Kim, Chim, Galli, and Wang}}]{feng}
\bibinfo{author}{\bibfnamefont{A.}~\bibnamefont{Splendiani}},
  \bibinfo{author}{\bibfnamefont{L.}~\bibnamefont{Sun}},
  \bibinfo{author}{\bibfnamefont{Y.}~\bibnamefont{Zhang}},
  \bibinfo{author}{\bibfnamefont{T.}~\bibnamefont{Li}},
  \bibinfo{author}{\bibfnamefont{J.}~\bibnamefont{Kim}},
  \bibinfo{author}{\bibfnamefont{C.-Y.} \bibnamefont{Chim}},
  \bibinfo{author}{\bibfnamefont{G.}~\bibnamefont{Galli}}, \bibnamefont{and}
  \bibinfo{author}{\bibfnamefont{F.}~\bibnamefont{Wang}},
  \bibinfo{journal}{Nano Lett.} \textbf{\bibinfo{volume}{10}},
  \bibinfo{pages}{1271} (\bibinfo{year}{2010}).

\bibitem[{\citenamefont{Wang et~al.}(2012)\citenamefont{Wang, Ruzicka, Kumar,
  Bellus, Chiu, and Zhao}}]{ultrafast1}
\bibinfo{author}{\bibfnamefont{R.}~\bibnamefont{Wang}},
  \bibinfo{author}{\bibfnamefont{B.~A.} \bibnamefont{Ruzicka}},
  \bibinfo{author}{\bibfnamefont{N.}~\bibnamefont{Kumar}},
  \bibinfo{author}{\bibfnamefont{M.~Z.} \bibnamefont{Bellus}},
  \bibinfo{author}{\bibfnamefont{H.-Y.} \bibnamefont{Chiu}}, \bibnamefont{and}
  \bibinfo{author}{\bibfnamefont{H.}~\bibnamefont{Zhao}},
  \bibinfo{journal}{Phys. Rev. B} \textbf{\bibinfo{volume}{86}},
  \bibinfo{pages}{045406} (\bibinfo{year}{2012}).

\bibitem[{\citenamefont{Xiao et~al.}(2012)\citenamefont{Xiao, Liu, Feng, Xu,
  and Yao}}]{spinvalley}
\bibinfo{author}{\bibfnamefont{D.}~\bibnamefont{Xiao}},
  \bibinfo{author}{\bibfnamefont{G.-B.} \bibnamefont{Liu}},
  \bibinfo{author}{\bibfnamefont{W.}~\bibnamefont{Feng}},
  \bibinfo{author}{\bibfnamefont{X.}~\bibnamefont{Xu}}, \bibnamefont{and}
  \bibinfo{author}{\bibfnamefont{W.}~\bibnamefont{Yao}},
  \bibinfo{journal}{Phys. Rev. Lett.} \textbf{\bibinfo{volume}{108}},
  \bibinfo{pages}{196802} (\bibinfo{year}{2012}).

\bibitem[{\citenamefont{Muniz and Sipe}(2015)}]{muniz}
\bibinfo{author}{\bibfnamefont{R.~A.} \bibnamefont{Muniz}} \bibnamefont{and}
  \bibinfo{author}{\bibfnamefont{J.~E.} \bibnamefont{Sipe}},
  \bibinfo{journal}{Phys. Rev. B} \textbf{\bibinfo{volume}{91}},
  \bibinfo{pages}{085404} (\bibinfo{year}{2015}).

\bibitem[{\citenamefont{Chernikov et~al.}(2014)\citenamefont{Chernikov,
  Berkelbach, Hill, Rigosi, Li, Aslan, Reichman, Hybertsen, and
  Heinz}}]{binding1}
\bibinfo{author}{\bibfnamefont{A.}~\bibnamefont{Chernikov}},
  \bibinfo{author}{\bibfnamefont{T.~C.} \bibnamefont{Berkelbach}},
  \bibinfo{author}{\bibfnamefont{H.~M.} \bibnamefont{Hill}},
  \bibinfo{author}{\bibfnamefont{A.}~\bibnamefont{Rigosi}},
  \bibinfo{author}{\bibfnamefont{Y.}~\bibnamefont{Li}},
  \bibinfo{author}{\bibfnamefont{O.~B.} \bibnamefont{Aslan}},
  \bibinfo{author}{\bibfnamefont{D.~R.} \bibnamefont{Reichman}},
  \bibinfo{author}{\bibfnamefont{M.~S.} \bibnamefont{Hybertsen}},
  \bibnamefont{and} \bibinfo{author}{\bibfnamefont{T.~F.} \bibnamefont{Heinz}},
  \bibinfo{journal}{Phys. Rev. Lett.} \textbf{\bibinfo{volume}{113}},
  \bibinfo{pages}{076802} (\bibinfo{year}{2014}).

\bibitem[{\citenamefont{He et~al.}(2014)\citenamefont{He, Kumar, Zhao, Wang,
  Mak, Zhao, and Shan}}]{binding2}
\bibinfo{author}{\bibfnamefont{K.}~\bibnamefont{He}},
  \bibinfo{author}{\bibfnamefont{N.}~\bibnamefont{Kumar}},
  \bibinfo{author}{\bibfnamefont{L.}~\bibnamefont{Zhao}},
  \bibinfo{author}{\bibfnamefont{Z.}~\bibnamefont{Wang}},
  \bibinfo{author}{\bibfnamefont{K.~F.} \bibnamefont{Mak}},
  \bibinfo{author}{\bibfnamefont{H.}~\bibnamefont{Zhao}}, \bibnamefont{and}
  \bibinfo{author}{\bibfnamefont{J.}~\bibnamefont{Shan}},
  \bibinfo{journal}{Phys. Rev. Lett.} \textbf{\bibinfo{volume}{113}},
  \bibinfo{pages}{026803} (\bibinfo{year}{2014}).

\bibitem[{\citenamefont{Kumar et~al.}(2013)\citenamefont{Kumar, Najmaei, Cui,
  Ceballos, Ajayan, Lou, and Zhao}}]{shgm}
\bibinfo{author}{\bibfnamefont{N.}~\bibnamefont{Kumar}},
  \bibinfo{author}{\bibfnamefont{S.}~\bibnamefont{Najmaei}},
  \bibinfo{author}{\bibfnamefont{Q.}~\bibnamefont{Cui}},
  \bibinfo{author}{\bibfnamefont{F.}~\bibnamefont{Ceballos}},
  \bibinfo{author}{\bibfnamefont{P.~M.} \bibnamefont{Ajayan}},
  \bibinfo{author}{\bibfnamefont{J.}~\bibnamefont{Lou}}, \bibnamefont{and}
  \bibinfo{author}{\bibfnamefont{H.}~\bibnamefont{Zhao}},
  \bibinfo{journal}{Phy. Rev. B} \textbf{\bibinfo{volume}{87}},
  \bibinfo{pages}{161403} (\bibinfo{year}{2013}).

\bibitem[{\citenamefont{Wang et~al.}(2013{\natexlab{a}})\citenamefont{Wang,
  Chien, Kumar, Kumar, Chiu, and Zhao}}]{thgm}
\bibinfo{author}{\bibfnamefont{R.}~\bibnamefont{Wang}},
  \bibinfo{author}{\bibfnamefont{H.-C.} \bibnamefont{Chien}},
  \bibinfo{author}{\bibfnamefont{J.}~\bibnamefont{Kumar}},
  \bibinfo{author}{\bibfnamefont{N.}~\bibnamefont{Kumar}},
  \bibinfo{author}{\bibfnamefont{H.-Y.} \bibnamefont{Chiu}}, \bibnamefont{and}
  \bibinfo{author}{\bibfnamefont{H.}~\bibnamefont{Zhao}}, \bibinfo{journal}{ACS
  Appl. Mater. Interfaces} \textbf{\bibinfo{volume}{6}}, \bibinfo{pages}{314}
  (\bibinfo{year}{2013}{\natexlab{a}}).

\bibitem[{\citenamefont{Torres-Torres et~al.}(2016)\citenamefont{Torres-Torres,
  Perea-L{\'o}pez, El{\'\i}as, Guti{\'e}rrez, Cullen, Berkdemir,
  L{\'o}pez-Ur{\'\i}as, Terrones, and Terrones}}]{thgw}
\bibinfo{author}{\bibfnamefont{C.}~\bibnamefont{Torres-Torres}},
  \bibinfo{author}{\bibfnamefont{N.}~\bibnamefont{Perea-L{\'o}pez}},
  \bibinfo{author}{\bibfnamefont{A.~L.} \bibnamefont{El{\'\i}as}},
  \bibinfo{author}{\bibfnamefont{H.~R.} \bibnamefont{Guti{\'e}rrez}},
  \bibinfo{author}{\bibfnamefont{D.~A.} \bibnamefont{Cullen}},
  \bibinfo{author}{\bibfnamefont{A.}~\bibnamefont{Berkdemir}},
  \bibinfo{author}{\bibfnamefont{F.}~\bibnamefont{L{\'o}pez-Ur{\'\i}as}},
  \bibinfo{author}{\bibfnamefont{H.}~\bibnamefont{Terrones}}, \bibnamefont{and}
  \bibinfo{author}{\bibfnamefont{M.}~\bibnamefont{Terrones}},
  \bibinfo{journal}{2D Mater.} \textbf{\bibinfo{volume}{3}},
  \bibinfo{pages}{021005} (\bibinfo{year}{2016}).

\bibitem[{\citenamefont{Wang et~al.}(2013{\natexlab{b}})\citenamefont{Wang,
  Wang, Fan, Lotya, O'Neill, Fox, Feng, Zhang, Jiang, and Zhao}}]{sat}
\bibinfo{author}{\bibfnamefont{K.}~\bibnamefont{Wang}},
  \bibinfo{author}{\bibfnamefont{J.}~\bibnamefont{Wang}},
  \bibinfo{author}{\bibfnamefont{J.}~\bibnamefont{Fan}},
  \bibinfo{author}{\bibfnamefont{M.}~\bibnamefont{Lotya}},
  \bibinfo{author}{\bibfnamefont{A.}~\bibnamefont{O'Neill}},
  \bibinfo{author}{\bibfnamefont{D.}~\bibnamefont{Fox}},
  \bibinfo{author}{\bibfnamefont{Y.}~\bibnamefont{Feng}},
  \bibinfo{author}{\bibfnamefont{X.}~\bibnamefont{Zhang}},
  \bibinfo{author}{\bibfnamefont{B.}~\bibnamefont{Jiang}}, \bibnamefont{and}
  \bibinfo{author}{\bibfnamefont{Q.}~\bibnamefont{Zhao}}, \bibinfo{journal}{ACS
  Nano} \textbf{\bibinfo{volume}{7}}, \bibinfo{pages}{9260}
  (\bibinfo{year}{2013}{\natexlab{b}}).

\bibitem[{\citenamefont{Geim and Grigorieva}(2013)}]{van}
\bibinfo{author}{\bibfnamefont{A.~K.} \bibnamefont{Geim}} \bibnamefont{and}
  \bibinfo{author}{\bibfnamefont{I.~V.} \bibnamefont{Grigorieva}},
  \bibinfo{journal}{Nature} \textbf{\bibinfo{volume}{499}},
  \bibinfo{pages}{419} (\bibinfo{year}{2013}).

\bibitem[{\citenamefont{Tongay et~al.}(2014)\citenamefont{Tongay, Sahin, Ko,
  Luce, Fan, Liu, Zhou, Huang, Ho, and Yan}}]{mono1}
\bibinfo{author}{\bibfnamefont{S.}~\bibnamefont{Tongay}},
  \bibinfo{author}{\bibfnamefont{H.}~\bibnamefont{Sahin}},
  \bibinfo{author}{\bibfnamefont{C.}~\bibnamefont{Ko}},
  \bibinfo{author}{\bibfnamefont{A.}~\bibnamefont{Luce}},
  \bibinfo{author}{\bibfnamefont{W.}~\bibnamefont{Fan}},
  \bibinfo{author}{\bibfnamefont{K.}~\bibnamefont{Liu}},
  \bibinfo{author}{\bibfnamefont{J.}~\bibnamefont{Zhou}},
  \bibinfo{author}{\bibfnamefont{Y.-S.} \bibnamefont{Huang}},
  \bibinfo{author}{\bibfnamefont{C.-H.} \bibnamefont{Ho}}, \bibnamefont{and}
  \bibinfo{author}{\bibfnamefont{J.}~\bibnamefont{Yan}}, \bibinfo{journal}{Nat.
  Commun.} \textbf{\bibinfo{volume}{5}} (\bibinfo{year}{2014}).

\bibitem[{\citenamefont{Cui and Zhao}(2015)}]{ballistic}
\bibinfo{author}{\bibfnamefont{Q.}~\bibnamefont{Cui}} \bibnamefont{and}
  \bibinfo{author}{\bibfnamefont{H.}~\bibnamefont{Zhao}}, \bibinfo{journal}{ACS
  Nano} \textbf{\bibinfo{volume}{9}}, \bibinfo{pages}{3935}
  (\bibinfo{year}{2015}).

\bibitem[{\citenamefont{Chenet et~al.}(2015)\citenamefont{Chenet, Aslan, Huang,
  Fan, van~der Zande, Heinz, and Hone}}]{ani2}
\bibinfo{author}{\bibfnamefont{D.~A.} \bibnamefont{Chenet}},
  \bibinfo{author}{\bibfnamefont{O.~B.} \bibnamefont{Aslan}},
  \bibinfo{author}{\bibfnamefont{P.~Y.} \bibnamefont{Huang}},
  \bibinfo{author}{\bibfnamefont{C.}~\bibnamefont{Fan}},
  \bibinfo{author}{\bibfnamefont{A.~M.} \bibnamefont{van~der Zande}},
  \bibinfo{author}{\bibfnamefont{T.~F.} \bibnamefont{Heinz}}, \bibnamefont{and}
  \bibinfo{author}{\bibfnamefont{J.~C.} \bibnamefont{Hone}},
  \bibinfo{journal}{Nano Lett.} \textbf{\bibinfo{volume}{15}},
  \bibinfo{pages}{5667} (\bibinfo{year}{2015}).

\bibitem[{\citenamefont{Cui et~al.}(2015)\citenamefont{Cui, He, Bellus,
  Mirzokarimov, Hofmann, Chiu, Antonik, He, Wang, and Zhao}}]{ani1}
\bibinfo{author}{\bibfnamefont{Q.}~\bibnamefont{Cui}},
  \bibinfo{author}{\bibfnamefont{J.}~\bibnamefont{He}},
  \bibinfo{author}{\bibfnamefont{M.~Z.} \bibnamefont{Bellus}},
  \bibinfo{author}{\bibfnamefont{M.}~\bibnamefont{Mirzokarimov}},
  \bibinfo{author}{\bibfnamefont{T.}~\bibnamefont{Hofmann}},
  \bibinfo{author}{\bibfnamefont{H.}~\bibnamefont{Chiu}},
  \bibinfo{author}{\bibfnamefont{M.}~\bibnamefont{Antonik}},
  \bibinfo{author}{\bibfnamefont{D.}~\bibnamefont{He}},
  \bibinfo{author}{\bibfnamefont{Y.}~\bibnamefont{Wang}}, \bibnamefont{and}
  \bibinfo{author}{\bibfnamefont{H.}~\bibnamefont{Zhao}},
  \bibinfo{journal}{Small} \textbf{\bibinfo{volume}{11}}, \bibinfo{pages}{5565}
  (\bibinfo{year}{2015}).

\bibitem[{\citenamefont{He et~al.}(2016)\citenamefont{He, Yan, Yin, Ye, Ye,
  Cheng, Li, and Lui}}]{stacking}
\bibinfo{author}{\bibfnamefont{R.}~\bibnamefont{He}},
  \bibinfo{author}{\bibfnamefont{J.-A.} \bibnamefont{Yan}},
  \bibinfo{author}{\bibfnamefont{Z.}~\bibnamefont{Yin}},
  \bibinfo{author}{\bibfnamefont{Z.}~\bibnamefont{Ye}},
  \bibinfo{author}{\bibfnamefont{G.}~\bibnamefont{Ye}},
  \bibinfo{author}{\bibfnamefont{J.}~\bibnamefont{Cheng}},
  \bibinfo{author}{\bibfnamefont{J.}~\bibnamefont{Li}}, \bibnamefont{and}
  \bibinfo{author}{\bibfnamefont{C.~H.} \bibnamefont{Lui}},
  \bibinfo{journal}{Nano Lett.}  (\bibinfo{year}{2016}).

\bibitem[{\citenamefont{Gao et~al.}(2016)\citenamefont{Gao, Li, Tan, Sun, Li,
  Idrobo, Singh, Lu, and Koratkar}}]{ap2}
\bibinfo{author}{\bibfnamefont{J.}~\bibnamefont{Gao}},
  \bibinfo{author}{\bibfnamefont{L.}~\bibnamefont{Li}},
  \bibinfo{author}{\bibfnamefont{J.}~\bibnamefont{Tan}},
  \bibinfo{author}{\bibfnamefont{H.}~\bibnamefont{Sun}},
  \bibinfo{author}{\bibfnamefont{B.}~\bibnamefont{Li}},
  \bibinfo{author}{\bibfnamefont{J.~C.} \bibnamefont{Idrobo}},
  \bibinfo{author}{\bibfnamefont{C.~V.} \bibnamefont{Singh}},
  \bibinfo{author}{\bibfnamefont{T.-M.} \bibnamefont{Lu}}, \bibnamefont{and}
  \bibinfo{author}{\bibfnamefont{N.}~\bibnamefont{Koratkar}},
  \bibinfo{journal}{Nano Lett.}  (\bibinfo{year}{2016}).

\bibitem[{\citenamefont{Zhang et~al.}(2016)\citenamefont{Zhang, Tan, Mendes,
  Sun, Chen, Kong, Xue, R{\"u}mmeli, Wu, and Chen}}]{ap3}
\bibinfo{author}{\bibfnamefont{Q.}~\bibnamefont{Zhang}},
  \bibinfo{author}{\bibfnamefont{S.}~\bibnamefont{Tan}},
  \bibinfo{author}{\bibfnamefont{R.~G.} \bibnamefont{Mendes}},
  \bibinfo{author}{\bibfnamefont{Z.}~\bibnamefont{Sun}},
  \bibinfo{author}{\bibfnamefont{Y.}~\bibnamefont{Chen}},
  \bibinfo{author}{\bibfnamefont{X.}~\bibnamefont{Kong}},
  \bibinfo{author}{\bibfnamefont{Y.}~\bibnamefont{Xue}},
  \bibinfo{author}{\bibfnamefont{M.~H.} \bibnamefont{R{\"u}mmeli}},
  \bibinfo{author}{\bibfnamefont{X.}~\bibnamefont{Wu}}, \bibnamefont{and}
  \bibinfo{author}{\bibfnamefont{S.}~\bibnamefont{Chen}},
  \bibinfo{journal}{Adv. Mater.}  (\bibinfo{year}{2016}).

\bibitem[{\citenamefont{Liu et~al.}(2015{\natexlab{a}})\citenamefont{Liu, Fu,
  Wang, Feng, Liu, Wan, Zhou, Wang, Shao, and Ho}}]{ap7}
\bibinfo{author}{\bibfnamefont{E.}~\bibnamefont{Liu}},
  \bibinfo{author}{\bibfnamefont{Y.}~\bibnamefont{Fu}},
  \bibinfo{author}{\bibfnamefont{Y.}~\bibnamefont{Wang}},
  \bibinfo{author}{\bibfnamefont{Y.}~\bibnamefont{Feng}},
  \bibinfo{author}{\bibfnamefont{H.}~\bibnamefont{Liu}},
  \bibinfo{author}{\bibfnamefont{X.}~\bibnamefont{Wan}},
  \bibinfo{author}{\bibfnamefont{W.}~\bibnamefont{Zhou}},
  \bibinfo{author}{\bibfnamefont{B.}~\bibnamefont{Wang}},
  \bibinfo{author}{\bibfnamefont{L.}~\bibnamefont{Shao}}, \bibnamefont{and}
  \bibinfo{author}{\bibfnamefont{C.-H.} \bibnamefont{Ho}},
  \bibinfo{journal}{Nat. Commun.} \textbf{\bibinfo{volume}{6}}
  (\bibinfo{year}{2015}{\natexlab{a}}).

\bibitem[{\citenamefont{Yang et~al.}(2016)\citenamefont{Yang, Kang, Yue, Coey,
  and Jiang}}]{ap1}
\bibinfo{author}{\bibfnamefont{S.}~\bibnamefont{Yang}},
  \bibinfo{author}{\bibfnamefont{J.}~\bibnamefont{Kang}},
  \bibinfo{author}{\bibfnamefont{Q.}~\bibnamefont{Yue}},
  \bibinfo{author}{\bibfnamefont{J.}~\bibnamefont{Coey}}, \bibnamefont{and}
  \bibinfo{author}{\bibfnamefont{C.}~\bibnamefont{Jiang}},
  \bibinfo{journal}{Adv. Mater. Interf.} \textbf{\bibinfo{volume}{3}}
  (\bibinfo{year}{2016}).

\bibitem[{\citenamefont{Zhang et~al.}(2015)\citenamefont{Zhang, Jin, Yuan,
  Wang, Zhang, Tang, Liu, Zhou, Hu, and Xiu}}]{ap4}
\bibinfo{author}{\bibfnamefont{E.}~\bibnamefont{Zhang}},
  \bibinfo{author}{\bibfnamefont{Y.}~\bibnamefont{Jin}},
  \bibinfo{author}{\bibfnamefont{X.}~\bibnamefont{Yuan}},
  \bibinfo{author}{\bibfnamefont{W.}~\bibnamefont{Wang}},
  \bibinfo{author}{\bibfnamefont{C.}~\bibnamefont{Zhang}},
  \bibinfo{author}{\bibfnamefont{L.}~\bibnamefont{Tang}},
  \bibinfo{author}{\bibfnamefont{S.}~\bibnamefont{Liu}},
  \bibinfo{author}{\bibfnamefont{P.}~\bibnamefont{Zhou}},
  \bibinfo{author}{\bibfnamefont{W.}~\bibnamefont{Hu}}, \bibnamefont{and}
  \bibinfo{author}{\bibfnamefont{F.}~\bibnamefont{Xiu}}, \bibinfo{journal}{Adv.
  Funct. Mater.} \textbf{\bibinfo{volume}{25}}, \bibinfo{pages}{4076}
  (\bibinfo{year}{2015}).

\bibitem[{\citenamefont{Najmzadeh et~al.}(2016)\citenamefont{Najmzadeh, Ko, Wu,
  Tongay, and Wu}}]{ap5}
\bibinfo{author}{\bibfnamefont{M.}~\bibnamefont{Najmzadeh}},
  \bibinfo{author}{\bibfnamefont{C.}~\bibnamefont{Ko}},
  \bibinfo{author}{\bibfnamefont{K.}~\bibnamefont{Wu}},
  \bibinfo{author}{\bibfnamefont{S.}~\bibnamefont{Tongay}}, \bibnamefont{and}
  \bibinfo{author}{\bibfnamefont{J.}~\bibnamefont{Wu}}, \bibinfo{journal}{Appl.
  Phys. Express} \textbf{\bibinfo{volume}{9}}, \bibinfo{pages}{055201}
  (\bibinfo{year}{2016}).

\bibitem[{\citenamefont{Liu et~al.}(2015{\natexlab{b}})\citenamefont{Liu,
  Zheng, He, Chaturvedi, He, Chow, Mion, Wang, Zhou, and Fu}}]{ap6}
\bibinfo{author}{\bibfnamefont{F.}~\bibnamefont{Liu}},
  \bibinfo{author}{\bibfnamefont{S.}~\bibnamefont{Zheng}},
  \bibinfo{author}{\bibfnamefont{X.}~\bibnamefont{He}},
  \bibinfo{author}{\bibfnamefont{A.}~\bibnamefont{Chaturvedi}},
  \bibinfo{author}{\bibfnamefont{J.}~\bibnamefont{He}},
  \bibinfo{author}{\bibfnamefont{W.~L.} \bibnamefont{Chow}},
  \bibinfo{author}{\bibfnamefont{T.~R.} \bibnamefont{Mion}},
  \bibinfo{author}{\bibfnamefont{X.}~\bibnamefont{Wang}},
  \bibinfo{author}{\bibfnamefont{J.}~\bibnamefont{Zhou}}, \bibnamefont{and}
  \bibinfo{author}{\bibfnamefont{Q.}~\bibnamefont{Fu}}, \bibinfo{journal}{Adv.
  Funct. Mater.} \textbf{\bibinfo{volume}{26}}, \bibinfo{pages}{1169}
  (\bibinfo{year}{2015}{\natexlab{b}}).

\bibitem[{\citenamefont{Murray et~al.}(1994)\citenamefont{Murray, Kelty,
  Chianelli, and Day}}]{ReS2s}
\bibinfo{author}{\bibfnamefont{H.~H.} \bibnamefont{Murray}},
  \bibinfo{author}{\bibfnamefont{S.~P.} \bibnamefont{Kelty}},
  \bibinfo{author}{\bibfnamefont{R.~R.} \bibnamefont{Chianelli}},
  \bibnamefont{and} \bibinfo{author}{\bibfnamefont{C.~S.} \bibnamefont{Day}},
  \bibinfo{journal}{Inorg. Chem.} \textbf{\bibinfo{volume}{33}},
  \bibinfo{pages}{4418} (\bibinfo{year}{1994}).

\bibitem[{\citenamefont{Moss et~al.}(1989)\citenamefont{Moss, van Driel, and
  Sipe}}]{sipe1}
\bibinfo{author}{\bibfnamefont{D.~J.} \bibnamefont{Moss}},
  \bibinfo{author}{\bibfnamefont{H.~M.} \bibnamefont{van Driel}},
  \bibnamefont{and} \bibinfo{author}{\bibfnamefont{J.~E.} \bibnamefont{Sipe}},
  \bibinfo{journal}{Opt. Lett.} \textbf{\bibinfo{volume}{14}},
  \bibinfo{pages}{57} (\bibinfo{year}{1989}).

\bibitem[{\citenamefont{Sipe et~al.}(1987)\citenamefont{Sipe, Moss, and
  Van~Driel}}]{sipe2}
\bibinfo{author}{\bibfnamefont{J.~E.} \bibnamefont{Sipe}},
  \bibinfo{author}{\bibfnamefont{D.~J.} \bibnamefont{Moss}}, \bibnamefont{and}
  \bibinfo{author}{\bibfnamefont{H.~M.} \bibnamefont{Van~Driel}},
  \bibinfo{journal}{Phys. Rev. B} \textbf{\bibinfo{volume}{35}},
  \bibinfo{pages}{1129} (\bibinfo{year}{1987}).

\bibitem[{\citenamefont{Castellanos-Gomez
  et~al.}(2014)\citenamefont{Castellanos-Gomez, Buscema, Molenaar, Singh,
  Janssen, van~der Zant, and Steele}}]{dry}
\bibinfo{author}{\bibfnamefont{A.}~\bibnamefont{Castellanos-Gomez}},
  \bibinfo{author}{\bibfnamefont{M.}~\bibnamefont{Buscema}},
  \bibinfo{author}{\bibfnamefont{R.}~\bibnamefont{Molenaar}},
  \bibinfo{author}{\bibfnamefont{V.}~\bibnamefont{Singh}},
  \bibinfo{author}{\bibfnamefont{L.}~\bibnamefont{Janssen}},
  \bibinfo{author}{\bibfnamefont{H.~S.} \bibnamefont{van~der Zant}},
  \bibnamefont{and} \bibinfo{author}{\bibfnamefont{G.~A.}
  \bibnamefont{Steele}}, \bibinfo{journal}{2D Mater.}
  \textbf{\bibinfo{volume}{1}}, \bibinfo{pages}{011002} (\bibinfo{year}{2014}).

\bibitem[{\citenamefont{Aslan et~al.}(2015)\citenamefont{Aslan, Chenet, van~der
  Zande, Hone, and Heinz}}]{tony}
\bibinfo{author}{\bibfnamefont{O.~B.} \bibnamefont{Aslan}},
  \bibinfo{author}{\bibfnamefont{D.~A.} \bibnamefont{Chenet}},
  \bibinfo{author}{\bibfnamefont{A.~M.} \bibnamefont{van~der Zande}},
  \bibinfo{author}{\bibfnamefont{J.~C.} \bibnamefont{Hone}}, \bibnamefont{and}
  \bibinfo{author}{\bibfnamefont{T.~F.} \bibnamefont{Heinz}},
  \bibinfo{journal}{ACS Photonics} \textbf{\bibinfo{volume}{3}},
  \bibinfo{pages}{96} (\bibinfo{year}{2015}).

%\bibitem[{\citenamefont{Boyd}(2003)}]{boyd}
%\bibinfo{author}{\bibfnamefont{R.~W.} \bibnamefont{Boyd}},
%  \emph{\bibinfo{title}{Nonlinear optics}} (\bibinfo{publisher}{Academic
%  press}, \bibinfo{year}{2003}).


\bibitem{sipe87} J. E. Sipe, J. Opt. Soc. Am. B {\bf 4}, 481-489 (1987).

\bibitem{supplement} See Supplemental Material at http://link.aps.org/supplemental  for details of the derivation of Eq. \ref{eq:chi3power}.

\bibitem[{\citenamefont{G{\"u}nter}(2012)}]{BK7}
\bibinfo{author}{\bibfnamefont{P.}~\bibnamefont{G{\"u}nter}},
  \emph{\bibinfo{title}{Nonlinear optical effects and materials}},
  vol.~\bibinfo{volume}{72} (\bibinfo{publisher}{Springer},
  \bibinfo{year}{2012}).

\end{thebibliography}

\end{document}